\documentclass[aps,prd,twocolumn,superscriptaddress,preprintnumbers,nofootinbib]{revtex4}

\preprint{NSF-KITP-14-011, FTPI-MINN-14/6, UMN-TH-3327/14}

\usepackage{epsfig}
\usepackage{amssymb}
\usepackage{amsmath}
\usepackage{amsfonts}
\usepackage{hyperref}

\newcommand{\be}{\begin{equation}}
\newcommand{\ee}{\end{equation}}
\newcommand{\bea}{\begin{eqnarray}}
\newcommand{\eea}{\end{eqnarray}}
\newcommand{\bi}{\begin{itemize}}
\newcommand{\ei}{\end{itemize}}
\newcommand{\ben}{\begin{enumerate}}
\newcommand{\een}{\end{enumerate}}

\def\gsim{\mathrel{\rlap{\lower4pt\hbox{\hskip1pt$\sim$}}
    \raise1pt\hbox{$>$}}}         %greater than or approx. symbol
\def\lsim{\mathrel{\rlap{\lower4pt\hbox{\hskip1pt$\sim$}}
    \raise1pt\hbox{$<$}}}         %less than or approx. symbol

\begin{document}

\title{
Remarks on the effect of bound states and threshold in  \boldmath{$g\!-\!2$}
}

\author{Kirill Melnikov}
\email{melnikov@pha.jhu.edu}
\affiliation{Department of Physics and Astronomy, Johns Hopkins 
University, Baltimore, USA}
\author{Arkady Vainshtein}
\email{vainshte@umn.edu}
\affiliation{William I. Fine Institute for Theoretical Physics and School of Physics and Astronomy,
University of Minnesota, Minneapolis, MN 54555, USA}
\affiliation{Kavli Institute for Theoretical Physics, University of California,Santa Barbara, CA 93106, USA}
\author{Mikhail  Voloshin}
\email{voloshin@umn.edu}
\affiliation{William I. Fine Institute for Theoretical Physics  and School of Physics and Astronomy, 
University of Minnesota, Minneapolis, MN 54555, USA}
\affiliation{Institute of Theoretical and Experimental Physics, Moscow 117218, Russia}

\begin{abstract}
Recently, the contribution 
% G.\,Mishima  and 
%M.\,Fael and M.\,Passera 
of positronium bound states to the electron anomalous magnetic moment was computed 
in Refs.\cite{mishima,fp}.
It was argued there
that this ${\cal O}(\alpha^5)$ contribution is missed if electron $g-2$ 
is calculated within conventional 
perturbative QED  and, as 
such, it must be added to the perturbative five-loop result. 
We show that this conclusion is flawed and that 
no additional contributions to $g-2$ are generated in QED beyond 
the perturbation theory.
\end{abstract}

\maketitle

Recently, the contribution of positronium bound states to electron anomalous 
magnetic moment was computed in Refs.\,\cite{mishima,fp}. The calculation 
proceeds  as follows.   Consider the contribution of the photon vacuum polarization 
in  QED ( with electrons and photons only)  to the electron $g-2$. We wish to compute 
this class of diagrams using 
dispersion representation for the vacuum polarization. The corresponding 
formula is well-known \cite{reviews}.  It reads 
\be
a_e({\rm vp})=\frac{g_{e}-2}{2} = \frac{\alpha}{\pi^2} \int \limits_{0}^{\infty} 
\frac{{\rm d} s }{s} \; {\rm Im} \,\Pi(s) \; K(s)\,,
\label{eq_ae}
\ee
where the kernel $K(s)$ is 
\be
K(s) = \int \limits_{0}^{1}\! {\rm d } x \,
\frac{x^2(1-x)}{x^2 + (1-x) s/m_e^2}\,.
\ee
The imaginary part of the vacuum polarization function 
$\Pi(s)$ can be represented  as a sum 
of   contributions of  positronium bound states  and the continuum contribution
\be
{\rm Im}\,\Pi(s)\! =\! 16\pi^{2}\sum \limits_{n=1}^{\infty} \frac{|\Psi_n(0)|^2}{M_{n}} \,\delta(s-M_n^2) + \theta(s-4m_e^2)\, C(s).
\label{eq_spd}
\ee
In that formula \cite{mishima,fp}, {$M_n\! =\! (2m_e +E_n)$}, {$E_n\! =\! - m_{e} \alpha^2/4n^2$},  
and $|\Psi_n(0)|^2 = m_{e}^3 \alpha^3/(8\pi n^3)$ are the parameters of the $S$-wave positronium bound 
states. 

It is well-known that bound-state contributions  
in Eq.\,(\ref{eq_spd}) can not be obtained at  any fixed order in 
perturbation theory in QED; rather a summation of infinite series of Feynman diagrams generated by exchanges 
of Coulomb photons is required.  Both Refs.\,\cite{mishima,fp} use this observation as an argument 
that the contribution of the positronium poles to $\Pi(s)$ 
is beyond the reach of conventional QED perturbation 
theory. They therefore suggest  that the shift in the anomalous magnetic moment \cite{mishima,fp} 
that one obtains 
by substituting the first term in Eq.\,(\ref{eq_spd}), representing the sum over bound states, into Eq.\,(\ref{eq_ae}) 
\be
a_{e}({\rm vp})^{\rm poles} = 
\frac{\alpha^5}{4\pi} \,\zeta(3) \left ( 8 \ln 2 - \frac{11}{2} \right )\,,
\label{eq_poles}
\ee
must be added to the five-loop result of Ref.\,\cite{kinoshita}. 

We would like to argue that this conclusion is wrong. Indeed, 
it is obvious that  this procedure can only be correct if 
the continuum in Eq.\,(\ref{eq_spd}) generates the same contribution to 
electron $g-2$ as what is obtained within 
 conventional perturbative QED, at least through  five-loops. We will show that this is {\it not} 
the case. In fact, the non-perturbative part of the pole contribution is completely canceled  by non-perturbative 
corrections to the continuum. This cancellation was actually discovered long ago in the 
framework of sum  rules, see Refs.\,\cite{cg,mv82,mv84}.

The non-perturbative nature of the positronium pole contributions can be understood as its  
non-analytic  dependence on the fine structure constant. Indeed, consider  
the case when the sign of $\alpha$ is reversed, so attraction is changed to repulsion. 
There is no bound states in this case and the pole contribution vanishes.
This implies that $\alpha$-expansion around  $\alpha =0$ can not be constructed for this contribution. 
To cover cases of both positive and negative $\alpha$,  we re-write the pole 
contribution as
\be
a_{e}({\rm vp})^{\rm poles} = 
\frac{\alpha^5+|\alpha|^{5}}{8\pi} \,\zeta(3) \left ( 8 \ln 2 - \frac{11}{2} \right )\,.
\label{eq_poles}
\ee
An appearance of $|\alpha|$ explicitly demonstrates non-analy-ticity.

We will now  show  that the non-analytical ${\cal O}(|\alpha|^5)$ term in Eq.(\ref{eq_poles}) 
is canceled  by the continuum contribution,
\be
a_e({\rm vp})^{\rm cont} = 
 \frac{\alpha}{\pi^2}\! \int \limits_{4m_e^2}^{\infty} \!
\frac{{\rm d} s }{s}\, C(s) K(s).
\label{eq_ae1}
\ee
We will focus on the lower integration region near the boundary $s = 4 m_e^2$ in Eq.\,(\ref{eq_ae1})
which is responsible for the non-analytical behavior.

Close to the electron-positron threshold, the imaginary part of the photon vacuum polarization 
function $\Pi(s)$ is known to all orders in $\alpha$. Indeed, for $s=(2m_e+E)^2$ with $E\ll m_{e}$
\be
\Pi(s) = \frac{2\pi\alpha}{m_e^2}\, G(0,0,E) + {\rm const}\,,
\label{eqps}
\ee
where the subtraction constant is included for the purpose of making the right-hand side finite, 
 and  $G(0,0,E)$ is the non-relativistic Green's function of the Coulomb problem $
G(\vec x, \vec y, E) = \langle \vec x | \left ( H - E \right )^{-1} | \vec y \rangle$.
Therefore, close to threshold, the continuum contribution is related to the imaginary 
part of the Coulomb Green's function which, for positive energies, is given by the Sommerfeld 
factor 
\be
C(s)  =\frac{2\pi\alpha}{m_e^2}\, {\rm Im}\,G(0,0,E)=
\frac{\pi  \alpha^2}{2} \frac{1}{1- e^{-\pi \alpha/\beta} }\,.
\label{eq_cs}
\ee
Here we have introduced the velocity $\beta = \sqrt{1-4m^2/s} \approx \sqrt{E/m}$\,.   Expanding this 
expression in $\alpha$, we obtain $\beta \to 0$ limit for $C(s)$ that is generated 
in fixed order perturbation theory of QED.
It reads 
\be
C(s)  = \frac{\alpha \beta}{2} \left ( 1 + \frac{\pi \alpha}{ 2 \beta}
 + \frac{\pi^2 \alpha^2}{12 \beta^2}  - \frac{\pi^4 \alpha^4}{720 \beta^4} +... \right ).
\ee
It is clear from this expression that the series do not converge in the threshold 
region where $\beta \sim \alpha$ and a proper computation of its contribution to $g-2$ 
requires summation of infinite series in $\alpha/\beta$.  By the criterion adopted 
in Refs.\,\cite{mishima,fp}, this 
is a contribution that is beyond any fixed-order perturbative calculation. 

To compute the contribution of this  $\beta \sim \alpha$ 
region to electron $g-2$, we write  ${\rm d} s \approx 8 m^2 \beta {\rm d \beta}$ and 
find  for the continuum part
\be
\begin{split}
& a_e({\rm vp})^{\rm cont} = 
\frac{\alpha^2}{\pi}  K(4 m_e^2) I(\alpha,\beta_0)\,, \\
& I(\alpha,\beta_0) =  \int \limits_{0}^{\beta_0}  {\rm d} \beta \,
\frac{\alpha\beta}{1-e^{-\pi \alpha/\beta}}\,.
\label{eq_b0}
\end{split}
\ee
The upper integration boundary $ \pi \alpha \ll \beta_0 \ll 1$ is introduced to make the 
threshold integral well defined. All contributions to $\Pi(s)$ coming from the region $\beta > \beta_0$ 
are obviously ``perturbative'', while the
``non-perturbative'' contribution appears from the integration over 
velocities $\beta \sim \alpha$. To isolate the contribution of this region, 
we subtract and add back the Taylor expansion of the integrand in Eq.\,(\ref{eq_b0}) 
at large $\beta$. We find 
\be
I(\alpha,\beta_0) = I_1(\alpha,\beta_0) + I_2(\alpha,\beta_0),
\ee
where
\be
\begin{split}
I_1 & = 
\int  \limits_{0}^{\beta_0}  {\rm d} \beta \beta\alpha \left [ 
\frac{\beta}{\pi \alpha} +\frac 1 2 + \frac{\pi\alpha}{12\beta}
 \right ]\;,\\
I_2 & = 
\int  \limits_{0}^{\beta_0}  {\rm d} \beta \beta \alpha \left [
\frac{1}{1-e^{-\pi \alpha/\beta}} - \frac{\beta}{\pi \alpha} - \frac 1 2- \frac{\pi\alpha}{12\beta}
 \right ].
\end{split} 
\ee
The integral $I_1$ can be computed explicitly; it is defined by the upper integration boundary 
and is ``perturbative''.  We do not consider it further. 

In the second integral $I_{2}$ 
due to its convergence at \mbox{$\beta \gg \pi \alpha$} the upper integration boundary 
$\beta_0$ can be set to infinity.  The integral $I_2$ is therefore 
``non-perturbative'': it receives contributions from the integration region 
\mbox{$\beta \sim \alpha \pi$} and it is not possible  
to compute it by perturbative expansion of the integrand in powers of $\alpha$. 

It is straigthforward to determine $I_{2}/\alpha^{3}$ by numerical integration.
To make analytical computation,  it is convenient to use the representation
\be
\begin{split} 
& \frac{1}{1\!-\!e^{-\pi \alpha/\beta}} - \frac{\beta}{\pi \alpha} 
- \frac 1 2 - \frac{\pi\alpha}{12\beta} \\
& =
 -\frac{\alpha}{2\pi\beta} \sum_{n=1}^{\infty}\frac{1}{n^{2}}\frac{(\alpha/2n)^{2}}{\beta^{2}+(\alpha/2n)^{2}}\,,
\end{split} 
\ee
which naturally appears in the Coulomb Green's function (see Eq.\,(\ref{green}) below).
Then, the integral $I_{2}$ takes the form
\be
I_{2}=-\frac{\alpha^{2}}{2\pi}\sum_{n=1}^{\infty}\frac{1}{n^{2}}\int  \limits_{0}^{\infty}  {\rm d} \beta\,\frac{(\alpha/2n)^{2}}{\beta^{2}+(\alpha/2n)^{2}}.
\ee
Note that the integral depends on $\alpha^{2}$ so it does not change sign when the sign of $\alpha$ is reversed.
The simple integration gives
\be
I_{2}=-\frac{|\alpha|^{3}}{8}\,\zeta(3)\,.
\ee

Finally, we  use   $K(4 m_e^2)  = 8 \ln(2) - 11/2\,$ and the above result for ${I}_2$ to 
derive the non-perturbative continuum contribution to $g-2$. 
We find 
\be
a_e({\rm vp})^{\rm cont,np} = -\frac{|\alpha|^5}{8 \pi} \,\zeta(3)\, \Big(8 \ln(2) - \frac{11}{2} \Big).
\label{eq_cont}
\ee
Adding  the pole contribution
Eq.~(\ref{eq_poles}) and the continuum contribution Eq.~(\ref{eq_cont}), 
we observe the  cancellation of the non-analytical  dependence on $|\alpha|$. 
The result reads
\be
a_e({\rm vp}) = \frac{\alpha^5}{8 \pi} \,\zeta(3)\, \Big(8 \ln(2) - \frac{11}{2} \Big).
\label{eqnptot}
\ee

We will now show that this analytic in $\alpha$ 
contribution can be obtained using conventional  perturbation theory, 
in spite of the fact that it appears to be coming from positronium poles, Eq.~(\ref{eq_poles}).
To see this, we focus on the threshold region where, as we already 
mentioned, the vacuum polarization contribution is proportional to Green's function of the 
Coulomb problem and where non-perturbative  modifications in the spectral density arise. 
 Ignoring changes of all functions that are smooth at threshold, 
we find 
\be
a_e({\rm vp})^{\rm thr}  = \frac{ 2\alpha^2 K(4 m_e^2)}{\pi m_e^3} 
\int {\rm d} E \; {\rm Im}\, G_E(0,0,E).
\label{eq_fe}
\ee
To calculate this integral, we note that the Green's function satisfies the dispersion 
relation 
\be
G_E(0,0,E) = \frac{1}{\pi}\!\int \limits_{E_1}^{\infty}\! {\rm d}E'\,\frac{{\rm Im}\, G(0,0,E')}{E' - E - i0}\,,
\ee
where \mbox{$E_1 = -m_e\alpha^2/4$} is the binding energy of the positronium ground state. Formally 
taking the limit 
$E \to -\infty$ in the above expression,  we obtain 
\be
\lim_{E \to -\infty} \pi E G_E(0,0,E) = - \int \limits_{E_1}^{\infty} {\rm d} E'\,
{\rm Im}\, G(0,0,E')\,.
\ee
Therefore, Eq.\,(\ref{eq_fe}) can be cast into the form where the integral of the spectral 
density is traded for the computation of Green's function of the Coulomb problem at 
large {\it negative} energy, far away from all the poles and singularities that are present 
in spectral density
\be
a_e({\rm vp})^{\rm thr}  \approx - \frac{2 \alpha^2  K(4 m_e^2)}{\pi m_e^3} \times\lim_{E \to -\infty} 
\left [  \pi E G_E(0,0,E) \right ].
\label{greenneg}
\ee
Green's function of a Coulomb problem at large negative energies can be calculated {\it 
perturbatively} without any reference to its spectral density and non-trivial effects 
there both in bound states and continuum, by iterating the equation
\be
G = G_0  -  G_0  V G 
\ee
in the Coulomb potential $V$.  This is exactly 
equivalent to what is done in conventional perturbative  computations, 
where all diagrams 
are  calculated
by performing the Wick rotation with subsequent integration
over the loop momenta.  In doing so, one maps the problem from a Minkowski 
one to an Euclidean one and cleanly avoids all singularities associated with 
multi-particle thresholds.  

Perturbative expansion of Green's function at large negative energies can be found from 
the following representation \cite{mv82,mv84}
\be
\begin{split} 
G(0,0,E) = & \frac{im_ek}{4\pi} - \frac{m_e^2 \alpha}{4\pi} \log ( -i k r_0) 
\\
&-\frac{m_{e}^{3}\alpha^{2}}{8\pi}\sum \limits_{n=1}^{\infty} \frac{1}{n^{2}}\frac{1}{(m_{e}\alpha/2n)+ik}\,,
\label{green}
\end{split} 
\ee
where $E = k^2/m_e,$ and $\log r_0$ is a subtraction constant that is absorbed in the 
constant term in Eq.(\ref{eqps}).
To compute the negative energy asymptotic, we use $k = i \kappa$, 
$\kappa > 0$.  Expanding Eq.\,(\ref{green}) in series in $\alpha$ and keeping track of the 
${\cal O}(\alpha^3)$ contribution, we obtain 
\be
G^{(3)}\Big(0,0,-\frac{\kappa^2}{m}\Big) =  \frac{m_e^4 \alpha^3 \zeta(3) }{16 \pi \kappa^2}\,.
\ee
Therefore, 
\be
\begin{split} 
& \lim_{E \to -\infty} 
\left [  \pi E G^{(3)}(0,0,E) \right ] = 
\\
&\lim_{\kappa \to \infty}  
 \left [ - \pi \frac{\kappa^2}{m} G^{(3)}  \right ] = - \frac{m_e^3 \alpha^3 \zeta(3) }{16}\,. 
\end{split} 
\ee
Using this result in Eq.\,(\ref{greenneg}), we obtain the  contribution 
shown in Eq.\,(\ref{eqnptot}). Since the above derivation relies solely on the properties 
of Green's function in the region where perturbative description is justified, we conclude that the non-perturbative 
effects cancel out in the sum of  positronium poles and continuum contributions.
The sum matches perturbative correction to $a_e({\rm vp})$ at order ${\cal O}(\alpha^5)$.

We note that absence of non-perturbative contributions can be also understood by 
regulating threshold singularities with  the photon mass. Indeed,  if we 
introduce the  non-vanishing photon mass $\lambda \ll m_e $ and 
keep it such that $\alpha m_e /\lambda \ll 1$, no bound states can appear in the spectral density. 
 However, the smooth limit $\lambda \to 0$ must exist in the each order in $\alpha$ which 
ensures that ``non-perturbative'' contributions are absent. Instead of the photon mass we can cut 
the Coulomb potential at distances $1/(m_{e}\alpha) \gg r \gg 1/m_{e}$ without changing the $g-2$ 
result (\ref{eqnptot}).

%One more line of argumentation is based on introduction of nonvanishing photon mass $\lambda$
%as an infrared regulator. 
%If in the $\alpha$ expansion we keep $\alpha m/\lambda \ll 1$ no bound states 
%appears whatsoever.

Finally, we note that the argumentation presented here is general and applies beyond  the case 
of anomalous magnetic moments. In fact, for any QED or QCD observable that requires the knowledge 
of any two-point function in the Eucledean region, perturbation theory gives complete description 
up to non-perturbative vacuum condensates parametrized by matrix elements of local operators
\cite{cg,mv82,mv84}. However, 
it never happens that a summation of infinite classes of Feynman diagrams enhanced at 
any threshold generates  additional effects beyond perturbation theory, in spite of highly non-trivial 
behavior of threshold spectral densities \cite{Smith:1994ev}.  
The misunderstanding of this fact, as illustrated by Refs.~\cite{mishima,fp} 
and a much earlier  discussion 
of how $t \bar t$ threshold effects may affect precision electroweak 
observables such as the $\rho$-parameter  \cite{kniehl2,kniehl1,yn}, appears to be quite common.  
We hope that the present  note will help to clarify it.

\noindent
{\bf Acknowledgments:} The work of K.M.  is partially supported by US NSF under grants PHY-1214000.
A.V. thanks for hospitality the Kavli Institute for Theoretical Physics where his research is supported 
 in part by the National Science Foundation under Grant No.\ NSF PHY11-25915. The work of M.V. is 
 supported in part by the DOE grant DE-FG02-94ER40823.

\end{document}